\documentclass[11pt]{article}
\usepackage{graphicx}
\usepackage{epsfig}
\usepackage{epsf,amsfonts,amssymb}
\usepackage{amsmath}

\newcommand{\be}{\begin{equation}}
\newcommand{\ee}{\end{equation}}
\newcommand{\bea}{\begin{eqnarray}}
\newcommand{\eea}{\end{eqnarray}}
\newcommand{\ba}{\begin{eqnarray}}
\newcommand{\ea}{\end{eqnarray}}

\newcommand{\eqn}[1]{(\ref{#1})}
\newcommand{\beq}{\begin{equation}}
\newcommand{\eeq}{\end{equation}}
\newcommand{\beqa}{\begin{eqnarray}}
\newcommand{\eeqa}{\end{eqnarray}}
\newcommand{\beqar}{\begin{eqnarray*}}
\newcommand{\eeqar}{\end{eqnarray*}}

\newcommand{\ie}{{\it i.e.,}\ }

\def\nc {N_\mt{c}}
\def\nf {N_\mt{f}}

\def\t6 {T_\mt{D6}}

\def\gym {g_\mt{YM}}

\newcommand{\te}{t_\mt{E}}

\newcommand{\lqcd}{\ensuremath{\Lambda_{\mt{QCD}}}}
\newcommand{\caln}{{\cal N}} 

\newcommand{\mt}[1]{\textrm{\tiny #1}}

\newcommand{\tdec}{T_\mt{dec}}

\newcommand{\ads}{\mbox{$AdS_5 \times S^5$ }}

\newcommand{\ra}{\rightarrow}

\begin{document}

\setlength{\unitlength}{1mm}

\thispagestyle{empty}
\vspace*{2cm}

\begin{center}
{\bf \Large String Theory and Quantum Chromodynamics\footnote{Lectures given at the RTN Winter School on ``Strings, Supergravity and Gauge Theories" at CERN on January 15-19, 2007.}}\\

\vspace*{1.4cm}

{\bf David Mateos}

\vspace*{0.2cm}

{\it Department of Physics, University of California, Santa Barbara, CA 93106-9530, USA}\\[.3em]


\vspace{2.8cm} {\bf ABSTRACT}
\end{center}

I review recent progress on the connection between string theory and quantum chromodynamics in the context of the gauge/gravity duality. Emphasis is placed on conciseness and conceptual aspects rather than on technical details. Topics covered include the large-$\nc$ limit of gauge theories, the gravitational description of gauge theory thermodynamics and hydrodynamics, and confinement/deconfinement thermal phase transitions.

\noindent

\vfill \setcounter{page}{0} \setcounter{footnote}{0}
\newpage

\tableofcontents

\section{Introduction}

Thirty-four years after the discovery of asymptotic freedom \cite{free}, Quantum Chromodynamics (QCD), the theory of the strong interactions between quarks and gluons, remains a challenge. There exist no analytic,  truly systematic methods with which to analyse its non-perturbative properties. Some of these, for example its thermodynamic properties, can be studied by means of the lattice formulation of QCD. However, other more dynamical ones, for example the transport properties of the quark-gluon plasma (QGP), are very hard to study on the lattice because of  the inherent Euclidean nature of this formulation. Even if in the future most features of QCD can be addressed on the lattice, as good as possible a theoretical understanding will still be desirable.

A long-standing hope is that a reformulation of QCD in terms a of a new set of string-like degrees of freedom would shed light on some of its mysterious properties. The purpose of these lectures is to review recent progress in the implementation of this idea in the context of the so-called `gauge/gravity' correspondence. 

We will begin by explaining why one ought to expect a stringy reformulation of QCD, or more generally of any gauge theory, to exist. 
We will then review one of the simplest examples of a gauge/gravity correspondence, namely the AdS/CFT correspondence between type IIB string theory on $AdS_5 \times S^5$ and 
four-dimensional $\caln = 4$ super Yang-Mills (SYM) theory. After that we will consider this correspondence at finite temperature, and we will see that this suffices to make contact with the physics of the deconfined QGP created in heavy ion collision experiments. 
In the following chapter we will consider a simple example of a confining theory with a gravity dual, and study the confinement/deconfinement phase transition that occurs as a function of the temperature. We will finish with a brief discussion of present limitations and challenges for the future.

The emphasis of this review is on conciseness and conceptual aspects rather than calculational details. It is also not exhaustive but rather the opposite, since the goal is to be able to discuss some of the most recent developments with as little technology as possible.

\section{Why QCD ought to have a string dual}
\label{why}
The expectation that it ought to be possible to reformulate QCD as a string theory can be motivated at different levels. Heuristically, the motivation comes from the fact that QCD is believed to contain string-like objects, namely the flux tubes between quark-antiquark pairs responsible for their confinement. Modelling these tubes by a string leads to so-called Regge behaviour, that is, the relation $M^2 \sim J$ between the mass and the angular momentum of the tube. The same behaviour is observed in the spectrum of mesons, \ie quark-antiquark bound states, in the real world. This argument, however, would not apply to non-confining gauge theories. 

A more precise motivation for the existence of a string dual of QCD, or more generally of any gauge theory, comes from consideration of the 't Hooft's large-$\nc$ limit 
\cite{thooft} (see \cite{wittenlarge} for a beautiful review). QCD is a gauge theory with gauge group $SU(3)$ and, because of dimensional transmutation, it possesses no expansion parameter. 't Hooft's idea is to consider a generalisation of QCD obtained by replacing the gauge group by $SU(\nc)$, to take the limit $\nc \rightarrow \infty$, and to perform an expansion in $1/\nc$. 

The degrees of freedom of this generalised theory are the gluon fields $A^i_{\mu j}$ and the quark fields $q^i_a$, where $i,j =1, \ldots , \nc$ and $a = 1, \ldots, \nf$, with $\nf$ the number of quark flavours. The number of independent gauge fields is $\nc^2 -1$ because of the fact the gauge group is $SU(\nc)$ and not $U(\nc)$, but since we will be working in the limit $\nc \rightarrow \infty$ we will ignore this difference. We will thus take the number of gluons to be $\sim \nc^2$. This is much larger than the number of quark degrees of freedom, $\nf \nc$, so we may expect (correctly) that the dynamics is dominated by the gluons in the large-$\nc$ limit. We will therefore start by studying the theory in this limit as if no quarks were present, and then examine the effect of their inclusion. 

\begin{figure}
\centering{\epsfxsize=8cm\epsfbox{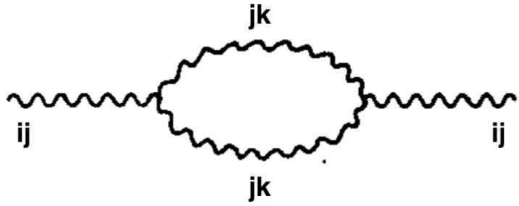}}
\caption{\small One-loop gluon self-energy  Feynman diagram.}
\label{gluon}
\end{figure}

To start with, consider the one-loop gluon self-energy Feynman diagram of fig.~\ref{gluon}. 
There are two vertices and one free colour index, so this scales as $\gym^2 \nc$. This means that for this diagram to possess a smooth limit in the limit 
$\nc \rightarrow \infty$, we must take at the same time $\gym \ra 0$ while keeping the so-called 't Hooft coupling $\lambda \equiv \gym^2 \nc$ fixed. This is equivalent to demanding that the confinement scale, $\lqcd$, remain fixed in the large-$\nc$ limit. This can be seen by noting that, with the scaling above, the one-loop $\beta$-function,
\be
\mu \frac{d}{d \mu} \gym^2 \propto -\nc \, \gym^4 \,,
\ee
becomes independent of $\nc$ when written in terms of $\lambda$:
\be
\mu \frac{d}{d \mu} \lambda \propto -\lambda^2 \,.
\ee

\begin{figure}
\centering{\epsfxsize=5cm\epsfbox{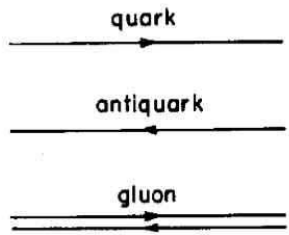}}
\caption{\small Double-line notation.}
\label{double}
\end{figure}

\begin{figure}
\centering{\epsfxsize=10cm\epsfbox{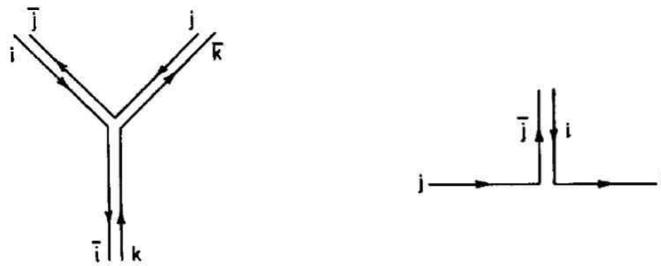}}
\caption{\small Vertices in double-line notation.}
\label{double-vertices}
\end{figure}

\begin{figure}
\centering{\epsfxsize=10cm\epsfbox{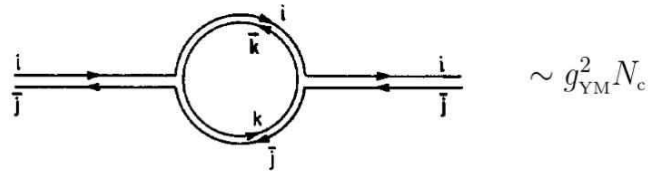}}
\caption{\small Gluon self-energy diagram in double-line notation.}
\label{double-vacuum}
\end{figure}

The determination of the $\nc$-scaling of Feynman diagrams is simplified by the so-called double-line notation. 
This consists of drawing the line associated to a gluon as a pair of lines associated to a quark and an anti-quark, as indicated in fig.~\ref{double}. In fig.~\ref{double-vertices} the three-gluon vertex (left) and the quark-antiquark-gluon vertex (right) are drawn in double-line notation. Fig.~\ref{double-vacuum} displays the one-loop gluon self-energy diagram of fig.~\ref{gluon} in this notation. We note that the factor of $\nc$ is associated to the `free' internal line carrying the index `$k$' in the figure.

\begin{figure}
\centering{\epsfxsize=12cm\epsfbox{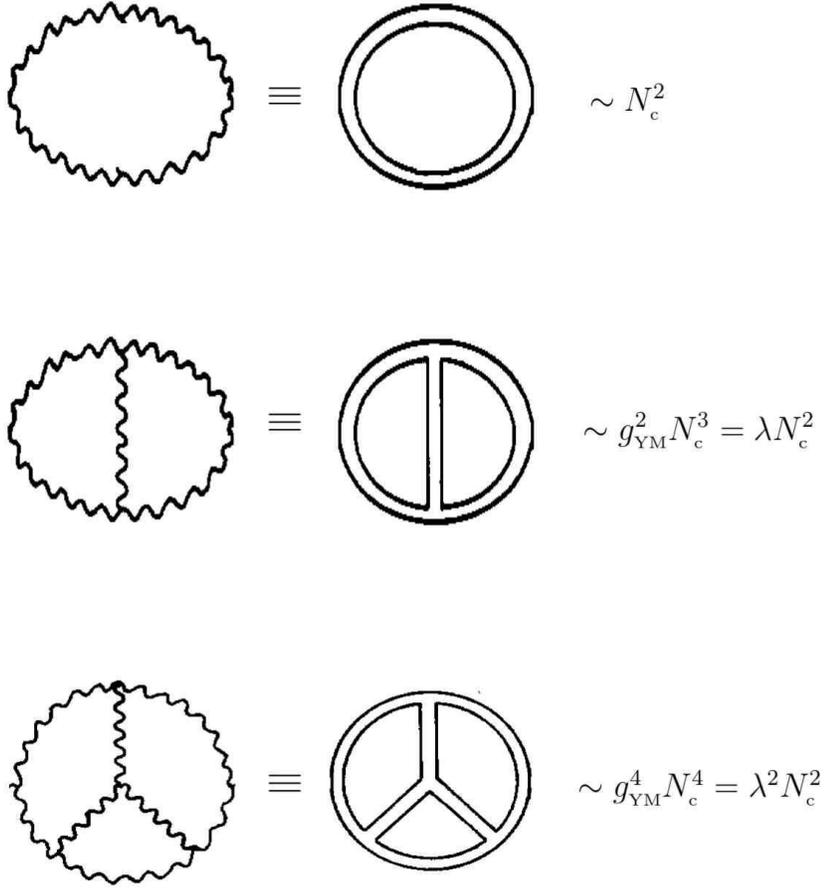}}
\caption{\small Some planar diagrams.}
\label{planar}
\end{figure}

\begin{figure}
\centering{\epsfxsize=12cm\epsfbox{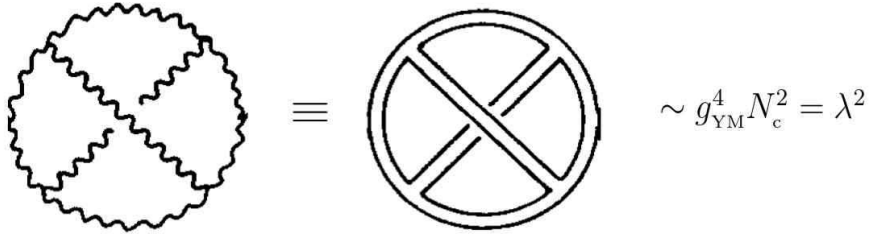}}
\caption{\small A non-planar diagram.}
\label{3loop}
\end{figure}

We will now see that Feynman diagrams naturally organise themselves in a double-series expansion in powers of $1/\nc$ and $\lambda$. For this purpose it suffices to consider a few vacuum diagrams (with no quarks for the time being). Some one-, two- and three-loop diagrams are shown in fig.~\ref{planar}. We see that they all scale with the same power of $\nc$ but a power of $\lambda^{\ell -1}$, with $\ell$ the number of loops. The $\nc^2$ scaling is not the same for all diagrams, however. 
For example, the three-loop diagram in fig.~\ref{3loop} scales as $\lambda^2$, and is thus suppressed with respect to those in fig.~\ref{planar} by a power of $1/\nc^2$; the reader is invited to draw other diagrams suppressed by higher powers of $1/\nc^2$. The difference between the diagrams in fig.~\ref{planar} and that of fig.~\ref{3loop} is that the former are planar, \ie can be `drawn without crossing lines', whereas the latter is not. We thus see that diagrams are classified by their topology, and that non-planar diagrams are suppressed in the large-$\nc$ limit. 

\begin{figure}
\centering{\epsfxsize=12cm\epsfbox{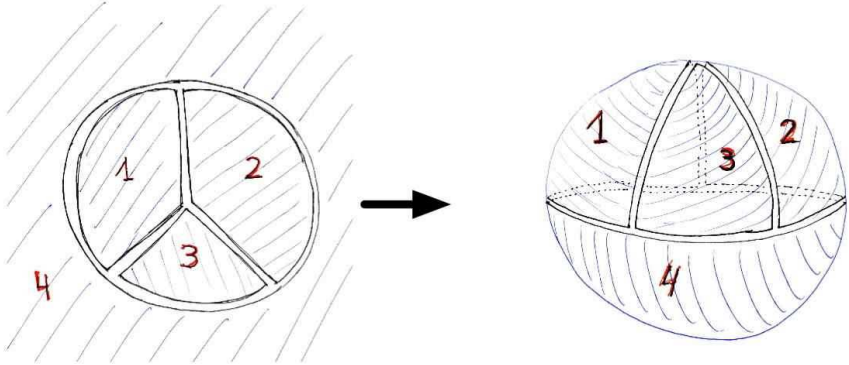}}
\caption{\small A Riemann surface associated to a planar diagram.}
\label{ssphere}
\end{figure}

\begin{figure}
\centering{\epsfxsize=12cm\epsfbox{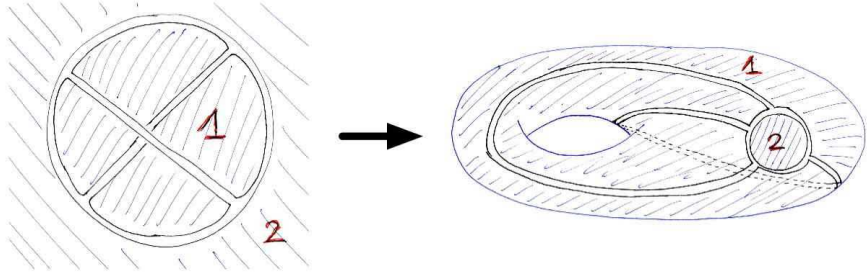}}
\caption{\small A Riemann surface associated to a non-planar diagram.}
\label{ttorus}
\end{figure}

The topological classification of diagrams, which leads to the connection with string theory, can be made more precise by associating a Riemann surface to each Feynman diagram, as follows. In double-line notation, each line in a Feynman diagram is a closed loop that we think of as the boundary of a two-dimensional surface or `face'. The Riemann surface is obtained by gluing together these faces along their boundaries as indicated by the Feynman diagram. In order to obtain a compact surface we add `the point at infinity' to the face associated to the external line in the diagram. This procedure is illustrated for a planar diagram in fig.~\ref{ssphere}, and for a non-planar diagram in fig.~\ref{ttorus}. 
In the first case we obtain a sphere, and the same is true for any planar diagram. In the second case we obtain a torus. It turns out that the power of $\nc$ associated to a given Feynman diagram is precisely $N_\mt{c}^\chi$, where $\chi$ is the Euler number of the corresponding Riemann surface. For a compact, orientable surface of genus $g$ with no boundaries we have $\chi = 2-2g$. Thus for the sphere $\chi=2$ and for the torus $\chi=0$. We therefore conclude that the expansion of any gauge theory amplitude in Feynman diagrams takes the form 
\be
{\cal A} = \sum_{g=0}^\infty \nc^\chi   \sum_{n=0}^\infty c_{g,n} \lambda^n \,,
\label{expansion}
\ee
where $c_{g,n}$ are constants. We recognise the first sum as the loop expansion in Riemann surfaces for a closed string theory with coupling constant $g_s \sim 1/\nc$. Note that the expansion parameter is therefore $1/\nc^2$. As we will see later, the second sum is associated to the so-called $\alpha'$-expansion in the string theory. 

\begin{figure}
\centering{\epsfxsize=12cm\epsfbox{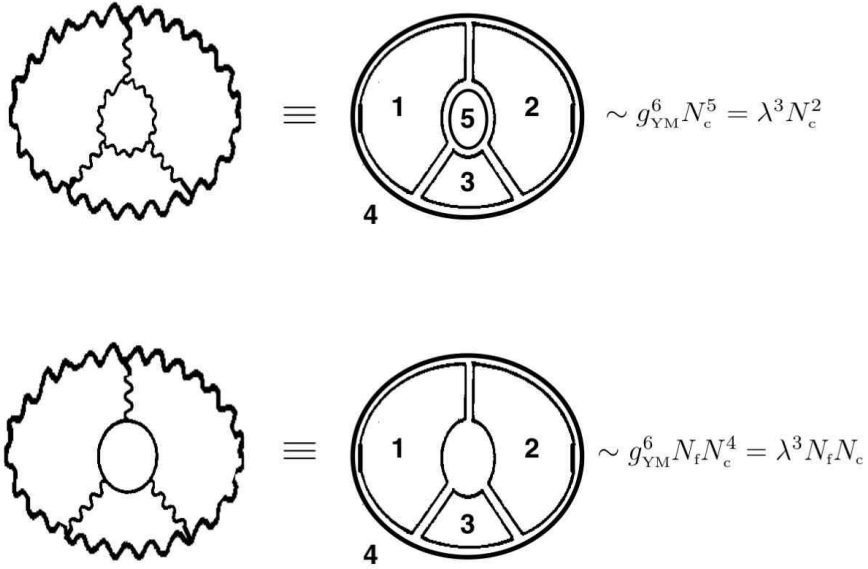}}
\caption{\small Two planar diagrams, one without quark loops (top diagram), and one with an internal quark loop (bottom diagram).}
\label{quark-supression}
\end{figure}

\begin{figure}
\centering{\epsfxsize=12cm\epsfbox{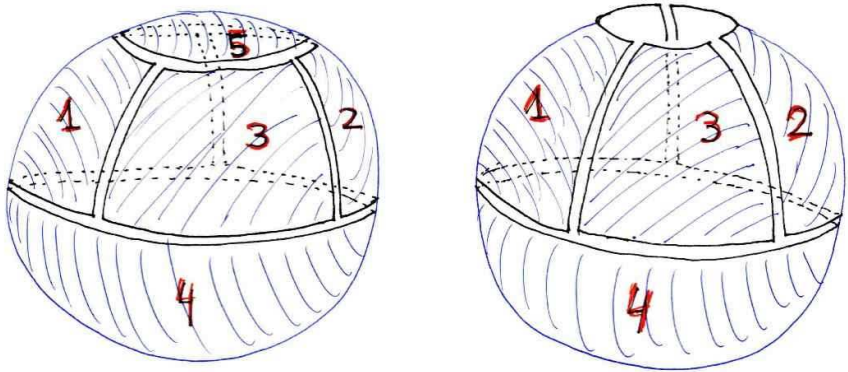}}
\caption{\small Riemann surfaces associated to the planar diagrams of fig.~\ref{quark-supression}. The surface on the left (right) corresponds to the top (bottom) diagram.}
\label{both}
\end{figure}

The above analysis holds for any gauge theory with Yang-Mills fields and possibly matter in the adjoint representation, since the latter is described by fields with two colour indices. 
In order to illustrate the effect of the inclusion of quarks, or more generally of matter in the fundamental representation, which is described by fields with only one colour index, consider the two diagrams in fig.~\ref{quark-supression}. 
The bottom diagram differs from the top diagram solely in the fact that a gluon internal loop has been replaced by a quark loop. This leads to one fewer free colour line and hence to one fewer power of $\nc$.  Since the flavour of the quark running in the loop must be summed over, it also leads to an additional power of $\nf$. Thus we conclude that internal quark loops are suppressed by powers of $\nf/\nc$ with respect to gluon loops. 
In terms of the Riemann surface associated to a Feynman diagram, the replacement of a gluon loop by a quark loop corresponds to the introduction of a boundary, as illustrated in fig.~\ref{both} for the diagrams of fig.~\ref{quark-supression}. 
The power of $\nc$ associated to the Feynman diagram is still $\nc^\chi$, but in the presence of $b$ boundaries the Euler number is $\chi=2-2g -b$. This means that in the large-$\nc$ expansion \eqn{expansion} we must also sum over the number of boundaries, and so we now recognise it as an expansion for a theory with both closed and open strings. The open strings are associated to the boundaries, and their coupling constant is 
$g_\mt{op} \sim \nf \, g_s^{1/2} = \nf/\nc$. 

The main conclusion of this section is therefore that the large-$\nc$ expansion of a gauge theory can be identified with the genus expansion of a string theory. Through this identification the planar limit of the gauge theory corresponds to the classical limit of the string theory. However, the analysis above does not tell us how to construct explicitly the string dual of a specific gauge theory. We will see that in some cases this can be `derived' by thinking about the physics of D-branes.

\section{The AdS/CFT correspondence}
\label{AdS/CFT}
In this section we will study one of the simplest examples of a gauge/gravity duality: The equivalence between type IIB string theory on $AdS_5 \times S^5$ and $\caln =4$ super Yang-Mills (SYM) theory on four-dimensional Minkowski space. Since this gauge theory is conformally invariant, this duality is an example of an AdS/CFT correspondence. We will see in later sections how non-AdS/non-conformal examples can be constructed.

\subsection{The decoupling limit}
\begin{figure}
\centering{\epsfxsize=3cm\epsfbox{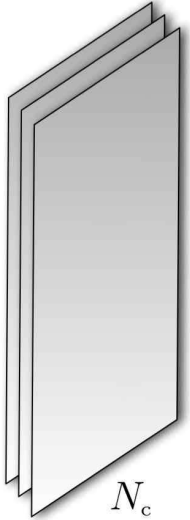}}
\caption{\small A set of $\nc$ D3-branes.}
\label{D3}
\end{figure}

To motivate the duality, let us consider the `ground-state' of type IIB string theory in the presence of $\nc$ D3-branes, as depicted in fig.~\ref{D3}. 
\begin{figure}
\centering{\epsfxsize=12cm\epsfbox{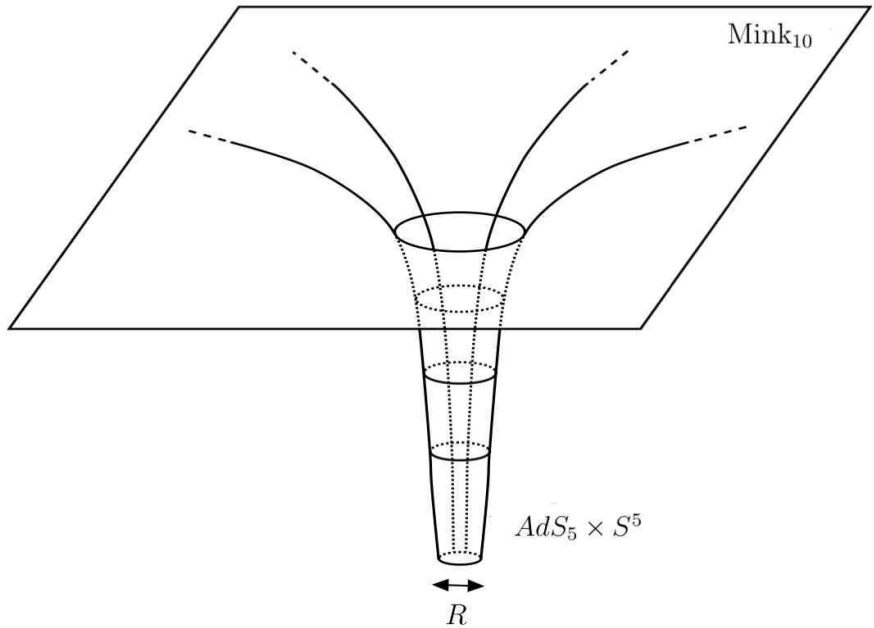}}
\caption{\small Spacetime around D3-branes.}
\label{AdS}
\end{figure}
Although the picture may suggest that the spacetime around the branes is flat, this is not true. D-branes carry mass and charge, and therefore curve the spacetime around them, as indicated in fig.~\ref{AdS}.

\begin{figure}
\centering{\epsfxsize=13cm\epsfbox{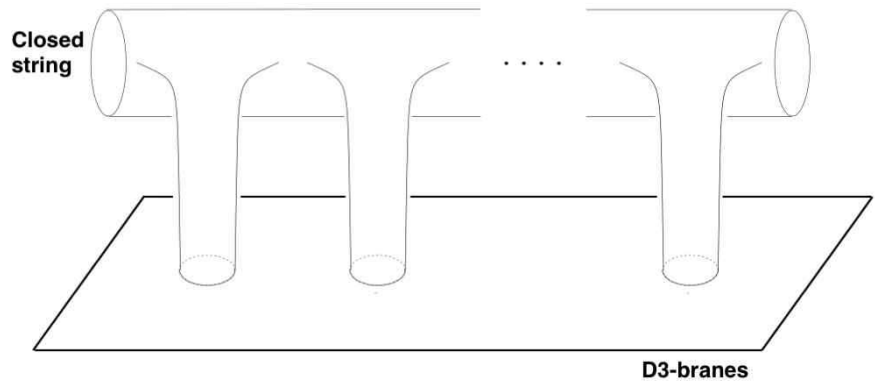}}
\vskip.2cm
\caption{\small Tadpole-like diagrams whose sum leads to an effective geometry for closed strings.}
\label{tadpole}
\end{figure}

Far away from the branes the spacetime is flat, ten-dimensional Minkowski space, whereas close to them a `throat' geometry of the form $AdS_5 \times S^5$ develops. 
Although this is not the way this spacetime is constructed in practice, conceptually it could be obtained by resumming an infinite number of tadpole-like diagrams with boundaries, of the form depicted in fig.~\ref{tadpole}, for a closed string propagating in the presence of the D3-branes. 

We would like to compare the gravitational radius $R$ of the D3-branes with the string length. On general grounds we expect (some power of) $R$ to be proportional to Newton's constant, to the number of D3-branes and to their tension. Newton's constant is given by
\be
16 \pi G = (2\pi)^7 g_s^2 \ell_s^8 \,,
\label{G}
\ee
with $\ell_s$ the string length. We note that it is proportional to $g_s^2$, and that in ten dimensions it has dimensions of length$^8$. The D3-branes are solitonic objects whose tension scales as an inverse power of the coupling, $T_\mt{D3} \sim 1/g_s \ell_s^4$.
It follows that the gravitational radius in string units must scale as $g_s \nc$. The precise relation turns out to be
\be
\frac{R^4}{\ell_s^4} = 4\pi g_s \nc \,.
\label{radius}
\ee
This means that if $g_s \nc \ll 1$ then the description suggested in fig.~\ref{D3} in terms of essentially zero-thickness objects in an otherwise flat spacetime is a good description. In this limit the the D3-branes are well described as a defect in spacetime, or more precisely as a boundary condition for open strings. Note that this conclusion relies crucially on the fact that the tension of D-branes scales as $1/g_s$ and not as $1/g_s^2$, as is typically the case for field theory solitons and as it would be the case for
NS5-branes in string theory.

In the opposite limit, $g_s \nc \gg 1$, the backreaction of the branes on a finite region of spacetime cannot be neglected, but fortunately in this case the description in terms of an effective geometry for closed strings becomes simple, since in this limit the size of the near-brane $AdS_5 \times S^5$ region becomes large in string units.

\begin{figure}
\centering{\epsfxsize=8cm\epsfbox{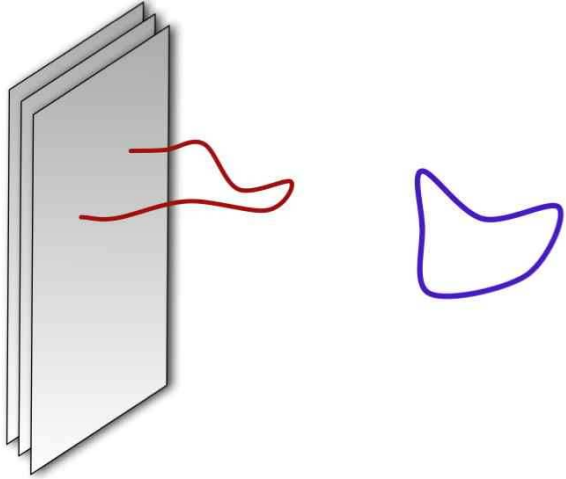}}
\vskip.2cm
\caption{\small Excitations of the system in the first description.}
\label{D3excitations}
\end{figure}

Now we are ready to motivate the AdS/CFT correspondence, by considering excitations around the ground-state in the two descriptions above and taking a low-energy or `decoupling' limit. In the first description the excitations of the system consist of open and closed strings, as displayed in fig.~\ref{D3excitations}, in interaction with each other. 
At low energies we may focus on the light degrees of freedom. Quantisation of the open strings leads to a spectrum consisting of a massless $\caln =4$ $SU(\nc)$ SYM multiplet plus a tower of massive string excitations. Since the open string endpoints are constrained to lie on the D3-branes, all these modes propagate in 3+1 flat dimensions -- the worldvolume of the branes. Similarly, quantisation of the closed strings leads to a massless graviton supermultiplet plus a tower of massive string modes, all of which propagate in flat ten-dimensional spacetime. The strength of interactions of closed string modes with each other is controlled by Newton's constant $G$, so the dimensionless coupling constant at an energy $E$ is $GE^8$. This vanishes at low energies and so in this limit closed strings become non-interacting, which is essentially the statement that gravity is infrared free. Interactions between closed and open strings are also controlled by the same parameter, since gravity couples universally to all forms of matter. Therefore at low energies closed strings decouple from open strings. In contrast, interactions between open strings are controlled by the $\caln =4$ SYM coupling constant in four dimensions, which is given by $\gym^2 \sim g_\mt{op}^2 \sim g_s$. 
Note that this relation is consistent with the fact that $\gym$ is dimensionless in four dimensions, and it can be derived, for example, by expanding the low-energy effective action for the D3-branes, the so-called Dirac-Born-Infeld (DBI) action:
\be
S_\mt{D3} \sim -T_\mt{D3} \int d^4 x \sqrt{-\det (\eta_{\mu\nu} + \alpha'^2 F_{\mu\nu} )} \sim  - \frac{1}{g_s} \int d^4 x \, F_{\mu\nu}^2  \,,
\ee
where $\alpha' = \ell_s^2$. We thus conclude that at low energies the first description of the system reduces to an interacting $\caln =4$ SYM theory in four dimensions plus free gravity in ten dimensions. 

\begin{figure}
\centering{\epsfxsize=12cm\epsfbox{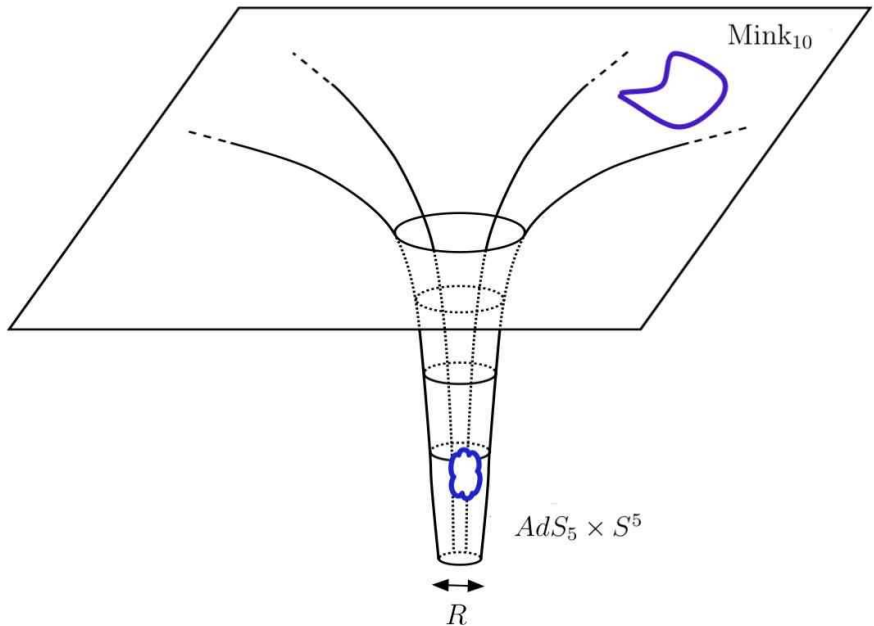}}
\vskip.2cm
\caption{\small Excitations of the system in the second description.}
\label{AdSexcitations}
\end{figure}

Let us now examine the same limit in the second description. In this case the low-energy limit consists of focusing on excitations that have arbitrarily low energy with respect to an observer in the asymptotically flat Minkowski region. As above, we now have two distinct sets of degrees of freedom, those propagating in the Minkowski region and those propagating in the throat -- see fig.~\ref{AdSexcitations}. 
In the Minkowski region the only modes that remain  are those of the massless ten-dimensional graviton supermultiplet. Moreover, at low energies these modes decouple from each other, since their interactions are governed by $GE^8$, as above. They also decouple from modes in the throat region, since at low energies the wave-length of these modes becomes much larger than the size of the throat. In the throat region, however, the whole tower of massive string excitations survives. This is because a mode in the throat must climb up a gravitational potential in order to reach the asymptotically flat region. Consequently, a closed string of arbitrarily high proper energy in the throat region may have an arbitrarily low energy as seen by an observer at asymptotic infinity, provided the string is located sufficiently deep down the throat. As we focus on lower and lower energies these modes become supported deeper and deeper in the throat as so they decouple from those in the asymptotic region.  We thus conclude that at low energies the second description of the system reduces to interacting closed strings in \ads plus free gravity in flat ten-dimensional spacetime. 

Comparing the results of the low-energy limits above it is reasonable to conjecture that four-dimensional $\caln =4$ $SU(\nc)$ SYM and type IIB string theory on $AdS_5 \times S^5$ are two apparently different descriptions of the same underlying physics \cite{maldacena}, and we will say that the two theories are `dual' to each other.

\subsection{Matching of parameters}
Let us examine more closely the parameters that enter the definition of each theory, and the map between them. The gauge theory is specified by the rank of the gauge group, $\nc$, and the 't Hooft coupling constant, $\lambda = \gym^2 \nc$. The string theory is determined by the string coupling constant $g_s$ and by the size of the $AdS_5$ and $S^5$ spaces. Both of these are maximally symmetric spaces which are completely specified by a single scale, their radius of curvature $R$. It turns out that the two spaces in the string solution sourced by D3-branes have equal radii. As we argued above, this is related to the parameters in the gauge theory through
\be
\frac{R^2}{\alpha'} \sim \sqrt{g_s \nc} \sim \sqrt{\lambda} \,.
\label{ff}
\ee
This means that the so-called $\alpha'$-expansion on the string side, which controls corrections associated to the finite size of the string as compared to the size of the spacetime it propagates in, corresponds to a strong-coupling, $1/\sqrt{\lambda}$ expansion in the gauge theory. 

It follows from \eqn{ff}  that a necessary condition in order for the particle or supergravity limit of the string theory to be a good approximation  we must have 
$\lambda \ra \infty$. Note, however, that this condition is not sufficient: It must be supplemented by the requirement that $g_s \ra 0$ (which then implies $\nc \ra \infty$) in order to ensure that additional degrees of freedom such as D-strings, whose tension scales as $1/g_s$, remain heavy. 

The string coupling is related to the gauge theory parameters through
\be
g_s \sim \gym^2 \sim \frac{\lambda}{\nc} \,,
\ee
which means that, for a fixed-size $AdS_5 \times S^5$ geometry 
(\ie for fixed $\lambda$), the string loop expansion corresponds precisely to the $1/\nc$ expansion in the gauge theory. Equivalently, one may note that the radius in Planck units is precisely 
\be
\frac{R^4}{\ell_p^4} \sim \frac{R^4}{\sqrt{G}} \sim \nc \,,
\ee
so quantum corrections on the string side are suppressed by powers of $1/\nc$.
In particular, the classical limit on the string side corresponds to the planar limit of the gauge theory.

\subsection{Matching of symmetries}
\begin{figure}
\centering{\epsfxsize=11cm\epsfbox{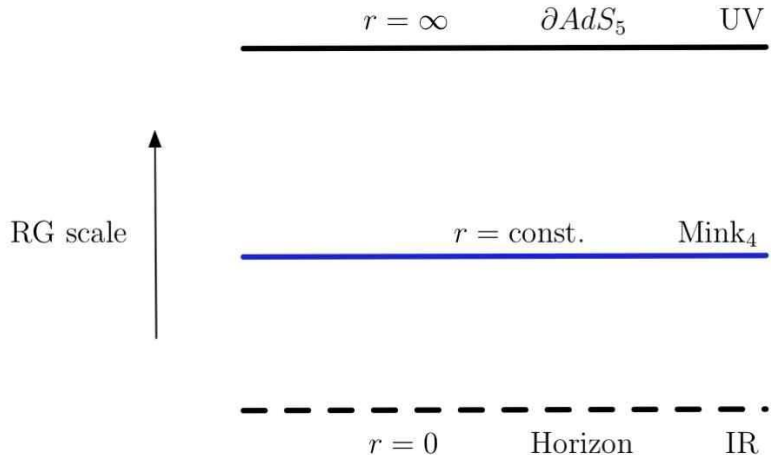}}
\caption{\small A geometric picture of $AdS_5$.}
\label{AdSpicture}
\end{figure}

The metric on $AdS_5$, in the so-called `Poincare patch', may be written as
\be
ds^2 = \frac{r^2}{R^2} \left( -dt^2 + dx_1^2 + dx_2^2 + dx_3^2 \right) + 
\frac{R^2}{r^2} dr^2 \,.
\label{AdSmetric}
\ee
The coordinates $x^\mu$ may be thought of as the coordinates along the worldvolume of the original D3-branes, and hence may be identified with the gauge theory coordinates. The coordinate $r$, and those on the $S^5$, span the directions transverse to the branes. As displayed in fig.~\ref{AdSpicture},
the coordinates used in \eqn{AdSmetric} provide a very simple geometric picture of $AdS_5$ as a foliation by constant-$r$ slices, each of which is isometric to four-dimensional Minkwoski spacetime. 
As $r \ra \infty$ we approach the so-called `boundary' of $AdS_5$. This is not a boundary in the topological but in the conformal sense of the word. Although this concept can be given a precise mathematical meaning, we will not need these details here. Since the norm of $\partial / \partial t$ vanishes at $r = 0$, we will refer to this surface as `the horizon'. Note, however, that the determinant of the induced three-metric on a constant-time slice vanishes at $r=0$, so this is not a finite-area horizon.

$\caln =4$ SYM is a conformal field theory (CFT). In particular, this means that it is invariant under the action of the dilatation operator 
\be
D: \,\,\,\, x^\mu \ra \Lambda x^\mu \,,
\label{D}
\ee
where $\Lambda$ is a constant. As one would expect, this transformation is also a symmetry on the gravity side: Indeed, the metric \eqn{AdSmetric} is invariant under \eqn{D} provided this is accompanied by the rescaling $r \ra r/\Lambda$. This means that short-distance physics in the gauge theory is associated to physics near the AdS boundary, whereas long-distance physics is associated to physics near the horizon. In other words, $r$ is identified with the renormalisation group (RG) scale in the gauge theory. Since a quantum field theory is defined by an ultraviolet (UV) fixed point and an RG flow,  one may think of the ${\cal N} =4$ gauge theory as residing at the boundary of $AdS_5$. The fact that $D$ acts on the $AdS_5$ metric as an exact isometry merely reflects the fact the RG flow is trivial for this gauge theory. For non-conformal theories with an UV fixed point $D$ is not an exact isometry of the dual geometry but only an asymptotic isometry. 

Let us examine more closely the matching of global symmetries on both sides of the correspondence. The $\caln =4$ SYM theory is invariant not only under dilatations but under $\mbox{Conf}(1,3) \times SO(6)$. The first factor is the conformal group of 
four-dimensional Minkowski space, which contains the Poincar\'e group, the dilatation symmetry generated by $D$, and four special conformal transformations whose generators we will denote by $K_\mu$. The second factor is the R-symmetry of the theory. In addition, the theory is invariant under sixteen ordinary or `Poincare' supersymmetries, the fermionic superpartners of the translation generators $P_\mu$, as well as under sixteen special conformal supersymmetries, the fermionic superpartners of the special conformal symmetry generators $K_\mu$. 

The string side of the correspondence is of course invariant under the group of  diffeomorphisms, which are gauge transformations. The subgroup of these consisting of large gauge transformations that leave the asymptotic (\ie near the boundary) form of the metric invariant is precisely $SO(2,4) \times SO(6)$. The first factor, which is isomorphic to $\mbox{Conf}(1,3)$,  corresponds to the isometry group of $AdS_5$, and the second one to the isometry group of $S^5$. As usual, large gauge transformations must be thought of as global symmetries, so we see that the bosonic global symmetry groups on both sides of the correspondence agree. An analogous statement can be made for the fermionic symmetries.  \ads is a maximally supersymmetric solution of type IIB string theory, and so it possesses thirty-two Killing spinors which generate fermionic isometries. These can be split into two groups that match those of the gauge theory. 

We therefore conclude that the global symmetries are the same on both sides of the duality. It is important to note, however, that on the gravity side the global symmetries arise as large gauge transformations. In this sense there is a correspondence between global symmetries in the gauge theory and gauge symmetries in the dual string theory. This is an important general feature of all known gauge/gravity dualities, to which we will return below after discussing the field/operator correspondence. It is also consistent with the general belief that the only conserved charges in a theory of quantum gravity are those associated to global symmetries that arise as large gauge transformations.

\subsection{The field/operator correspondence} 
So far we have not provided a precise prescription for the map between observables in the two theories. The technical details will not be needed in these lectures, but we will now sketch the main idea \cite{main}. This can be motivated by recalling that the SYM coupling constant $\gym^2$ is identified with the string coupling constant $g_s$. In string theory this is given by $g_s=e^{\Phi_\infty}$, where $\Phi_\infty$ is the value of the dilaton at  the AdS boundary. This suggests that deforming the gauge theory by changing the value of a coupling constant corresponds to changing the value of a string field at $\partial$AdS. More generally, one may imagine deforming the gauge theory action as
\be
S \ra S + \int d^4 x \, \phi(x) {\cal O} (x) \,,
\ee
where ${\cal O} (x)$ is a gauge-invariant, local operator and $\phi(x)$ is a possibly point-dependent coupling, namely a source. It is then reasonable to expect that to each such possible operator there corresponds a dual string field $\Phi(x,r)$ such that its value at the AdS boundary may be regarded as a source for the above operator, \ie we identify $\phi = \left. \Phi \right|_{\partial AdS}$. For example, the dilaton field is dual to (roughly) the operator $\mbox{Tr} F^2$. A natural conjecture is then that the partition functions of the two theories agree upon this identification, namely that
\be
Z_\mt{CFT} \left[ \phi \right] = Z_\mt{string} \left[ \left. \Phi \right|_{\partial AdS} \right] \,.
\label{Z}
\ee
The left-hand side encodes all the physical information in the gauge theory, since it allows the calculation of correlation functions of arbitrary gauge-invariant operators.\footnote{The prescription may be extended to non-local operators, such as Wilson loops 
\cite{wilson}.} The right-hand side is in general not easy to compute, but it simplifies dramatically in the large-$\nc$, large-$\lambda$ limit, in which it reduces to 
\be
Z_\mt{string} \simeq e^{-S_\mt{sugra}} \,, 
\label{S}
\ee
where $S_\mt{sugra}$ is the on-shell supergravity action.
 
An especially important set of operators in a gauge theory are conserved currents associated to global symmetries. Given the correspondence between these and gauge  symmetries on the string side, we expect the field dual to a conserved current $J^\mu$  to be a gauge field $A_\mu$. This is indeed true, and is consistent with the fact that the coupling
\be
\int d^4 x \, A_\mu(x) J^\mu (x) 
\ee
is invariant under gauge transformations $\delta A_\mu = \partial_\mu f$ by virtue of the fact that $\partial_\mu J^\mu =0$.

A particular set of currents that are conserved in any translationally invariant theory are those in the energy-momentum tensor operator $T_{\mu \nu}$. This must couple to a symmetric, spin-two gauge field, namely to a graviton, in the form
\be
\int d^4 x \, g^{\mu\nu} (x) T_{\mu\nu} (x) \,.
\ee
Thus we reach the general conclusion that the dual of a translationally invariant gauge theory must involve dynamical gravity.

\subsection{Remarks} 
Let us close this section with a few general remarks. First, the AdS/CFT correspondence described here is not proven, but it has passed an large number of tests. In these lectures we will assume that it holds in its strongest form, \ie for all values of $\lambda$ and $\nc$. Second, the correspondence is a deep statement about the equivalence of two a priori completely different theories. The CFT is just a somewhat exotic example of a system based on rules we are familiar with, those of quantum field theory in four-dimensional Minkowski spacetime. However, string theory is a quantum theory of gravity, so the correspondence implies, for example, that the CFT knows about a sum over all possible geometries with AdS boundary conditions. These may include geometries with non-equivalent topologies, with or without black holes, etc. Finally, the AdS/CFT correspondence is perhaps our most concrete implementation of the holographic principle \cite{holo}, since a theory of quantum gravity (in this case string theory) in a given spacetime is stated to be equivalent to a theory residing on its boundary.

\section{Finite temperature and RHIC physics}
\label{finite}
In this section we will modify the correspondence above by considering finite temperature physics. One motivation for this is as follows. At zero temperature 
$\caln =4$ SYM and QCD are very different theories. QCD is a confining theory with a dynamically generated scale $\lqcd \simeq 200$ MeV, whereas $\caln =4$ SYM is a conformal theory with no scales. Moreover,  $\caln =4$ SYM is highly supersymmetric, whereas QCD is not. However, at a temperature $\tdec \simeq 170$ MeV, QCD is believed to undergo a cross-over to a deconfined phase referred to as the `quark-gluon plasma' (QGP) phase. Since any finite temperature breaks both the supersymmetry and the conformal invariance of the $\caln =4$ SYM theory, one may hope that some properties of the $\caln =4$ plasma may be shared by the QCD plasma.

\subsection{Finite-temperature AdS/CFT}
The framework of the previous section is easily modified to introduce a finite temperature $T$. We obtained the zero-temperature correspondence by taking a decoupling limit of extremal D3-branes, which saturate the BPS bound $M=|Q|$. Adding temperature means adding energy but no charge to the system, so it is natural to take a decoupling limit for non-extremal D3-branes. It turns out that the net effect of this is solely to modify the AdS part of the metric, replacing \eqn{AdSmetric} by 
\be
ds^2 = \frac{r^2}{R^2} \left( - f dt^2 + dx_1^2 + dx_2^2 + dx_3^2 \right) + 
\frac{R^2}{r^2 f} dr^2 \,,
\label{AdSfinite}
\ee
where
\be
f(r) = 1 - \frac{r_0^4}{r^4} 
\ee
and $r_0$ is a constant with dimensions of length related to the temperature. 

The metric \eqn{AdSfinite} coincides with \eqn{AdSmetric} for large values of $r$. Recalling that $r$ is related to the energy scale in the gauge theory, we see that this is merely the statement that the ultraviolet physics in unaffected by the temperature, as we would expect. However, the infrared behaviour of the metric \eqn{AdSfinite} is very different from that of \eqn{AdSmetric}. The metric \eqn{AdSfinite} possesses a regular, finite-area horizon at $r=r_0$. The Hawking temperature of this horizon is interpreted as the temperature of the dual CFT. The simplest way to calculate it is to demand that  the Euclidean continuation of the metric \eqn{AdSfinite}, 
\be
ds_\mt{E}^2 = \frac{r^2}{R^2} \left( f d\te^2 + dx_1^2 + dx_2^2 + dx_3^2 \right) + 
\frac{R^2}{r^2 f} dr^2 \,,
\label{AdSeuclidean}
\ee
obtained as usual by the replacement $t \ra i \te$, be regular. Since the Euclidean time direction shrinks to zero size at $r=r_0$, we must require that $\te$ be periodically identified with appropriate period $\beta$, \ie $\te \sim \te +\beta$. A simple calculation shows that 
\be
\beta = \frac{\pi R^2}{r_0} \,.
\label{beta}
\ee
The period $\beta$ of the Euclidean time circle is then interpreted as the inverse temperature, $\beta = 1/T$. The reason for this is that, at finite temperature $T$, one is interested in calculating the partition function $\mbox{Tr} \, e^{-\beta H}$, where $H$ is the Hamiltonian of the theory. In a path integral formulation, the trace may be implemented by periodically identifying the Euclidean time with period $\beta$.

\subsection{Thermodynamics: Entropy density}
We are now ready to perform a simple but important calculation, namely that of the entropy density of large-$\nc$ $\caln =4$ SYM at strong coupling  \cite{entropy}. We do not know how to compute this in the gauge theory, but in the limit 
$\nc, \lambda \ra \infty$ we can use the supergravity description. In this description the entropy is just the Bekenstein-Hawking entropy, $S_\mt{BH}=A/4G$, proportional to the area of the horizon in the metric \eqn{AdSfinite}, or, more precisely, in the ten-dimensional metric consisting of the direct product of 
\eqn{AdSfinite} with a five-sphere of radius $R$. The horizon lies at $r=r_0$ and 
$t=\mbox{const.}$, and has `area'
\be
A = \int d^3 x \, d^5\Omega \, \sqrt{g} \,.
\ee
The determinant of the metric factorises into the determinant of the metric on $S^5$ times $r_0^3/R^3$, where the latter factor is just the determinant of the three-metric on a $r=r_0, t=\mbox{const.}$ slice in \eqn{AdSfinite}. Integrating we obtain $A = a V_3$, where 
\be
a = \frac{r_0^3}{R^3}  \times \pi^3 R^5 \,,
\ee
$V_3 = \int d^3 x$ is the (infinite) volume in the 123-directions and $\pi^3$ is the volume of a unit five-sphere. Using \eqn{G}, \eqn{radius} and \eqn{beta}
we can express the entropy density per unit volume in the 123-directions in terms of gauge theory parameters as
\be
s_\mt{BH}=  \frac{a}{4 G} = \frac{\pi^2}{2} \nc^2 T^3 \,.
\label{d3entropy}
\ee
The $\nc$ and the temperature dependence of this result could have been anticipated. The former follows from the fact that the number of degrees of freedom in an 
$SU(\nc)$ gauge theory grows as $\nc^2$, whereas the latter follows from dimensional analysis, since the temperature is the only scale in the $\caln=4$ theory. What is truly remarkable about the result above is that it shows that the entropy density attains a finite value in the limit of infinite coupling, $\lambda \ra \infty$. Moreover, this result differs from the result at zero coupling merely by a (famous) factor of $3/4$. Indeed, the Bekenstein-Hawking entropy density above is related to that of a free gas of 
$\caln=4$ particles through 
\be
s_\mt{BH} = \frac{3}{4} s_\mt{free} \,.
\ee
The significance of this result will become clearer when we discuss the thermodynamics of QCD.

\subsection{Hydrodynamics: Shear viscosity}
Above we used the AdS/CFT correspondence to calculate one example of a thermodynamic quantity at strong coupling, namely the entropy density. In this section we will sketch the calculation of an important hydrodynamic coefficient, the shear viscosity of the $\caln=4$ plasma \cite{PSS,from}. 

\begin{figure}
\centering{\epsfxsize=7cm\epsfbox{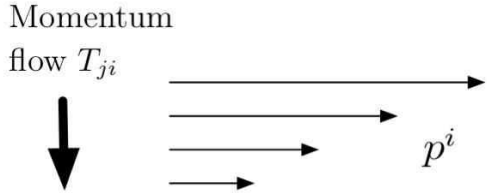}}
\vskip.2cm
\caption{\small Trasnport of $i$-momentum in the $j$-direction.}
\label{viscosity}
\end{figure}

While thermodynamics describes static properties of a system in perfect thermal equilibrium, hydrodynamics is the effective theory that describes long-wavelength, small-amplitude perturbations around thermal equilibrium. As any effective theory, hydrodynamics requires the knowledge of a few parameters, such as transport coefficients, that must determined with the microscopic theory. One important such coefficient is the shear viscosity, $\eta$, which measures the diffusion of momentum in a given direction $i$ along an orthogonal direction $j$. This momentum flow is measured by the component $T_{ij}$ of the energy-momentum tensor, and, in the linear-response approximation, it is proportional to the gradient of $i$-momentum density in the $j$-direction, $\nabla_j T_{0i}$ (see fig.~\ref{viscosity}). 
The shear viscosity is thus defined (roughly) through the constituent relation 
$T_{ij} \sim  -\eta \nabla_j T_{0i}$. 

One word of caution about the reader's possible intuition is in order. One may have thought that weakly coupled systems, such as a gas of weakly interacting particles, have low viscosities. In fact, the opposite is true. The reason is that the viscosity is proportional to the mean free path of the particles,  since the longer this is the more efficiently momentum is transported between regions with different momenta. 
Particles in weakly interacting gases have long free paths, leading to large viscosities, whereas particles in strongly coupled liquids have short free paths, leading to small viscosities.  

In a quantum field theory, the viscosity can be calculated by means of a Kubo-type formula as 
\be
\eta = - \lim_{\omega \ra 0} \, \frac{1}{\omega} \, \mbox{Im} \, 
G^\mt{R}_{ij,ij} \left( \omega, \vec{q} =0 \right) \,,
\label{kubo}
\ee
where $G^\mt{R}$ is the retarded correlator of two energy-momentum tensors
\be
G^\mt{R}_{\mu\nu, \alpha \beta} \left( q \right) = 
-i \int d^4 x \, e^{-iqx} \theta(t) 
\langle \left[ T_{\mu\nu}(x), T_{\alpha \beta} (0) \right] \rangle \,,
\label{retarded}
\ee
and $q=(\omega, \vec{q})$.
The fact that the viscosity is given by a formula like \eqn{kubo} is not surprising. The correlator $G^\mt{R}$ measures the response at a point $x$ to a perturbation at a point $x=0$. Evaluating the zero-momentum, low-frequency limit of this correlator corresponds to the long-wavelength limit of hydrodynamics. The imaginary part is associated to a diffusion-like process, in this case of momentum density.

With the formula \eqn{kubo} in hand it is a simple problem to calculate the viscosity of the $\caln =4$ plasma at large-$\nc$ and strong coupling using the supergravity description. However, this is a slightly technical calculation that requires, among other things, the details of the real-time prescription for the calculation of correlators 
\cite{mink,realtime}. Therefore here we will only sketch the necessary steps and refer the reader to the original references for the details. 

In order to extract the viscosity using the formula \eqn{kubo} one must calculate the retarded correlator \eqn{retarded}. This can be obtained by taking two functional derivatives of the gauge theory generating functional with respect to an appropriate source that couples to the energy-momentum tensor. We saw in the previous section that the AdS/CFT duality identifies the generating functional of the gauge theory with that of the string theory, see eqn.~\eqn{Z}, which in the supergravity approximation reduces to eqn.~\eqn{S}. Moreover, the energy-momentum tensor $T_{\mu\nu}$ of the gauge theory is dual to the metric $g_{\mu\nu}$ on the string side. Therefore the desired correlator is schematically given by
\be
\langle T T \rangle \sim \left. \frac{\delta^2}{\delta h^2} S_\mt{sugra} 
\left[ g+h \right] \right|_{h=0} \,,
\label{TT}
\ee
where $S_\mt{sugra}$ is the on-shell supergravity action, $g$ is the metric 
\eqn{AdSfinite} and $h$ is an infinitessimal metric perturbation. Since one is only interested in a second derivative, it suffices to consider the supergravity action expanded to quadratic order in this perturbation, which leads to a linear equation of motion. Once this is solved, the result can be substituted back into the action and the derivative in \eqn{TT} evaluated. The result for the viscosity is
\be
\eta = \frac{\pi}{8} \nc^2 T^3 \,.
\ee

\subsection{The viscosity/entropy ratio}
The hydrodynamic behaviour of a system is better characterised by the ratio of its shear viscosity to its entropy density, $\eta/s$, rather than by $\eta$ itself, since this ratio is a measure of the viscosity per degree of freedom. From the results above it follows that for $\caln=4$ SYM
\be
\frac{\eta}{s} = \frac{1}{4\pi} \,.
\label{ratio}
\ee
This result is important because both explicit calculations \cite{explicit1, explicit2} and general arguments \cite{general1, general2} have shown that it is a universal property of large-$\nc$, strongly coupled, finite temperature gauge theories with a gravity dual. 
These include theories in different numbers of dimensions, with or without a chemical potential, with or without fundamental matter, etc. This universality does not hold for other transport coefficients. Presumably, the reason for the universality of $\eta/s$ is that both the entropy density and the shear viscosity are related to universal properties of black hole horizons. Using the AdS/CFT prescription, one can show that under very general conditions (which however do not include theories with chemical potentials) the shear viscosity is given by \cite{KSS}
\be
\eta = \frac{\sigma_\mt{abs} (\omega \ra 0)}{16 \pi G} \,,
\ee
where $\sigma_\mt{abs} (\omega \ra 0)$ is the zero-fequency limit of the absorption cross-section of the black hole for a minimally coupled scalar. Again under very general conditions one can show that this is precisely equal to the area of the black hole horizon, $\sigma_\mt{abs} (\omega \ra 0) = a$ \cite{abs}. Since the entropy density is $s=a/4G$ we obtain \eqn{ratio}.

\begin{figure}
\centering{\epsfxsize=7cm\epsfbox{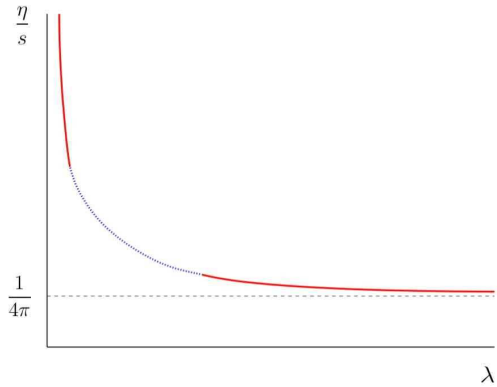}}
\caption{\small Simplest possibility for the behaviour of $\eta/s$ as a function of the coupling.}
\label{eta-over-s}
\end{figure}

An important feature of the ratio \eqn{ratio} is that it is very small compared to that of most substances in Nature. For example, $\eta/s \simeq 380 /4\pi$ for water, after which hydrodynamics is named, whereas $\eta /s \simeq 9/ 4\pi$ for liquid helium. For a weakly coupled quantum field theory the leading-order result is
\be
\frac{\eta}{s} = \frac{A}{\lambda^2 \log (B / \sqrt{\lambda})} \,,
\label{pert}
\ee
where $A,B$ are constants; for QCD with $\nf=3$ one finds 
$(A,B) \simeq (46,4)$ \cite{AMY}, whereas for $\caln=4$ SYM the result is $(A,B) \simeq (6,2)$ \cite{HJM}. Thus the ratio is very large at weak coupling. Corrections to the entropy density \cite{entropycorrection} and the shear viscosity 
\cite{viscositycorrection} of $\caln=4$ SYM associated to going away from the strict 
$\lambda \ra \infty$ limit have been calculated, with the result that the ratio \eqn{ratio} is modified to 
\be
\frac{\eta}{s} = \frac{1}{4\pi} \left( 1 + \frac{k}{\lambda^{3/2}} + \cdots \right) \,,
\ee
where $k$ is positive constant. The simplest possibility is therefore that, at large-$\nc$, the coupling dependence of the ratio for $\caln=4$ SYM is that of a monotonically decreasing function, as sketched in fig.~\ref{eta-over-s}. 
One type of $1/\nc$ corrections associated to the effect of fundamental matter has also been calculated and shown to modify both $\eta$ and $s$ but to leave the ratio equal to $1/4\pi$ \cite{us,us2}.

The above evidence supports the conjecture of \cite{KSS} that 
\be
\frac{\eta}{s} \geq \frac{1}{4\pi} 
\label{bound}
\ee
may be a universal bound obeyed by all physical systems, in particular by relativistic quantum field theories at finite temperature. We will return to this in the next subsection.

\subsection{Relation to QCD}
At low temperatures the quarks and gluons of QCD are confined and the physical degrees of freedom consist of colour-singlet hadrons. In this phase thermodynamic quantities scale as $\nc^0$. At a critical temperature $\tdec$, a rapid crossover into a new phase occurs (in the large-$\nc$ limit this is actually a first-order phase transition). In the new phase the quarks and gluons are deconfined, and so this phase is referred to as the 
QGP phase. In this phase thermodynamic quantities scale as $\nc^2$. At sufficiently high temperatures the theory becomes weakly coupled and a description directly in terms of these degrees of freedom is appropriate. However, we will see that this is not necessarily the case  at temperatures above but close to $\tdec$. 

\begin{figure}
\centering{\epsfxsize=10cm\epsfbox{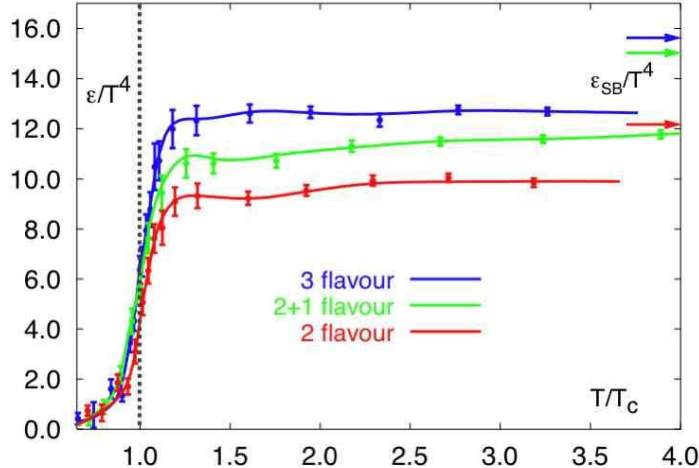}}
\caption{\small Lattice calculation of the energy density in QCD.}
\label{crossover}
\end{figure}

The crossover above manifests itself, for example, in the behaviour of thermodynamic quantities. Fig.~\ref{crossover} displays the lattice calculation of one such quantity, the energy density, as a function of $T/\tdec$ \cite{energy}. Lattice calculations of the deconfinement temperature $\tdec$ range between 151 MeV and 192 MeV \cite{crossover}. 
The rapid increase of the energy density around $T \sim \tdec$ reflects the liberation or deconfinement of the quark and gluon degrees of freedom. A remarkable property of this plot is the fact that the energy density over $T^4$ stays practically constant between  
$\tdec$ and at least $4\tdec$. Moreover, the value of the energy density in this plateau is around $80\%$ of the Stefan-Boltzman value for a free gas of quarks and gluons. This result was interpreted by part of the community as an indication that the QGP at temperatures just above $\tdec$ is already a weakly coupled gas of quarks and gluons. After all, the correction to the energy density associated to interactions amounts `only' to a $20\%$ difference with respect to the free result. This belief then led to the expectation that the ratio of viscosity to entropy density for the QGP should be large, $\eta/s \gg 1$, since as we saw above weakly coupled systems have large viscosities -- see eqn.~\eqn{pert}.

\begin{figure}
\centering{\epsfxsize=10cm\epsfbox{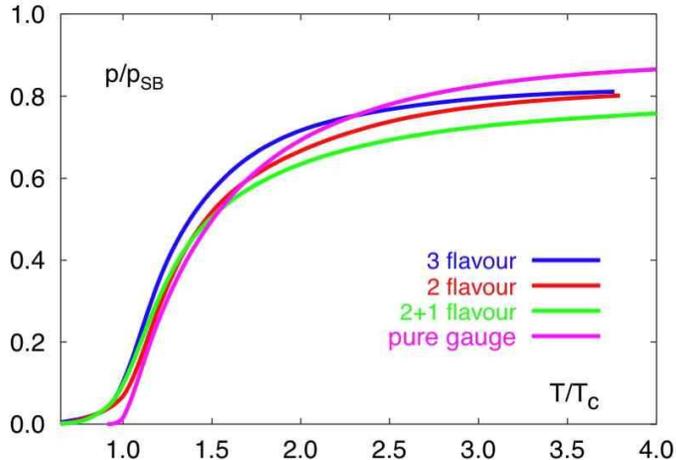}}
\caption{\small Lattice calculation of the pressure in QCD.}
\label{pressure}
\end{figure}

However, the interpretation above may seem (rightfully so) counter-intuitive. First of all, other thermodynamic quantities such as the pressure, for example, do not reach a constant value just above $\tdec$, but instead rise much more slowly, as shown in fig.~\ref{pressure} \cite{energy}. 
More conceptually, one may ask what the basis is to expect that a theory that is sufficiently strongly coupled in the infrared to produce confinement becomes weakly coupled immediately after it deconfines. Of course, because of asymptotic freedom, we should certainly expect the QGP to become weakly coupled at asymptotically high temperatures, but there is in principle no reason to expect that regime to extend all the way down to $T \gtrsim \tdec$. Finally, the fact that some thermodynamic quantities take a value very close to that of the free theory does not mean that the coupling is weak. We even know a counter-example to this: As we saw above, the difference in entropy densities of $\caln =4$ SYM at zero and infinite coupling is just a factor of 
$3/4$! The cunning reader will not fail to notice that $3/4 = 75 \% \simeq 80 \%$. 
In hindsight, it is tempting to reinterpret the lattice result for the energy density in QCD as suggesting that the QCD plasma at $T \gtrsim \tdec$ is strongly coupled. 

The $\caln=4$ theory also illustrates perfectly a more general phenomenon, namely the fact that the thermodynamics of a theory may be very similar at weak and strong coupling, and yet its hydrodynamics, characterised by transport coefficients, may be radically different. For example, the ratio $\eta/s$ diverges as $\lambda \ra 0$, but it approaches $1/4\pi$ as $\lambda \ra \infty$. As we explained above, the latter limit of this ratio seems to be a universal property of all non-Abelian plasmas for which a gravity dual with which to calculate the ratio at strong coupling is available. This suggests that this is a robust property that does not depend on certain details of the plasma, such as the precise field content, the presence of a chemical potential, etc. One may thus conjecture that, as far as this ratio is concerned, `all non-Abelian, strongly coupled plasmas look alike'. If true, this would suggest that if the QCD plasma at $T \gtrsim \tdec$ is strongly coupled, then the ratio $\eta/s$ should be close to $1/4\pi$.

Unlike that of thermodynamic quantities, the lattice calculation of dynamical properties of the QCD plasma, such as its shear viscosity, is problematic due to the inherent Euclidean nature of the lattice formulation.\footnote{See however \cite{meyer} for a recent attempt.} 
We therefore have virtually no methods at present to calculate the ratio $\eta/s$ for QCD at temperatures just above deconfinement. However, this ratio can be extracted from heavy ion collision experiments such as those performed at the Relativistic Heavy Ion Collider (RHIC) in Brookhaven, NY.

Heavy ions such as gold nuclei are collided at RHIC at centre-of-mass energies around 200 GeV/nucleon. There is evidence that a thermally equilibrated QGP is formed in such collisions, with temperatures that range approximately between $\tdec$ and 
$2 \tdec$ \cite{shuryak}. Some of this evidence comes from the successful reproduction of experimental data by hydrodynamic simulations, which assume local thermal equilibrium. There is also evidence (for example from back-to-back jet suppression with respect to $pp$-collisions) that the plasma behaves as a strongly coupled liquid. Finally, the comparison between hydrodynamic simulations and experimental data suggests that the ratio $\eta/s$ is close to $1/4\pi$. Initial estimates  
\cite{teaney} suggested a value slightly above the bound \eqn{bound}, 
\be
\frac{\eta}{s} \simeq (2-4) \times \frac{1}{4\pi} \,, 
\ee
whereas more recent ones \cite{recent} favour a lower value  
\be
\frac{\eta}{s} \simeq \frac{1}{2} \times \frac{1}{4\pi} \,.
\ee
Important assumptions that are hard to verify independently, such as the nature of the initial state, enter these hydrodynamic simulations, so the above results must be taken with caution. Thus at present it is unclear whether or not the bound \eqn{bound} may be violated by the QCD plasma. However, it does seem clear that the ratio $\eta/s$ is close to $1/4\pi$, as suggested by the gauge/gravity correspondence.

\section{Confinement/deconfinement phase transitions}
\label{phase}
In the previous sections we have concentrated on what is perhaps the simplest example of a gauge/gravity duality, the equivalence between type IIB string theory on \ads and four-dimensional $\caln=4$ SYM. The latter theory is very different from QCD in many respects, one of the most important ones being that it does not exhibit confinement. In this section we will study a simple example of a confining theory with a gravity dual, as well as the gravitational analogue of the confinement/deconfinement phase transition.

\subsection{A confining theory from D4-branes}
One way to construct a confining gauge theory with a gravity dual, due to Witten \cite{wittenD4}, is to start with $\nc$ D4-branes, instead of with $\nc$ D3-branes as we did in section \ref{AdS/CFT}. The gauge theory on the D4-branes is a maximally supersymmetric, $SU(\nc)$ SYM theory in five dimensions whose field content consists, in addition to the gluons, of scalars and fermions in the adjoint representation.
In order to obtain a four-dimenional theory, consider compactifying one spacelike direction of the D4-branes on a circle $S^1_L$ of length $L$. Since we would like to break supersymmety, we impose antiperiodic boundary conditions around the circle for the fermions. This projects out their zero-mode, so from the four-dimensional viewpoint they acquire a tree-level mass of order $M = 1/L$. Through quantum effects, a mass is also generated for the scalars. The only degrees of freedom that remain massless in four dimensions are the zero modes of the gauge fields around the circle, since a mass for these modes is forbidden by gauge invariance. Thus at energies $E \ll M$ the theory ought to reduce to pure $SU(\nc)$ gluodynamics, which we expect to confine at some dynamically generated scale $\lqcd$. We also expect that a confinement/deconfinement phase transition should occur at a temperature $\tdec \sim \lqcd$. 

Before we proceed to construct the gravitational description of the D4-brane theory above, let us note that we would like to study this theory in the regime in which 
$\lqcd \ll M$, so that the dynamics we are interested in is not contaminated by the presence of additional fields at the scale $M$. We will see in this section that, unfortunately, this limit cannot be described in the dual string theory using solely supergravity. We will discuss the implications of this in the last section. 

The gravity solution dual to the theory on the D4-branes is easily constructed along the lines of previous sections. One starts with the solution for near-extremal D4-branes and takes an appropriate decoupling limit. The result is the ten-dimensional metric
\be
ds^2 = \left( \frac{r}{R} \right)^{3/2} \left( -f dt^2 + dx_{(3)}^2 + dy^2 \right) +
\left( \frac{R}{r} \right)^{3/2} \frac{dr^2}{f} + R^{3/2} r^{1/2} ds^2 \left( S^4 \right) \,,
\label{D4metric}
\ee
where 
\be
f(r) = 1- \frac{r_0^3}{r^3} \,.
\ee
The last term on the right-hand side is the round metric on a unit four-sphere. The coordinates on this sphere, together with the radial coordinate $r$, span the five-dimensional space transverse to the D4-branes.  The coordinates $\{t, x_{(3)}, y\}$ span the five-dimensional worldvolume of the D4-branes and are identified with the gauge theory coordinates. In particular, this means that we must periodically identify 
$y \sim y+L$, with $L=1/M$.  

Unlike the metric on $AdS_5 \times S^5$, the metric \eqn{D4metric}  does not factorise into a direct product of some spacetime times a sphere, since the radius of the $S^4$ is not constant. This is important for the calculation of many quantities, but it will play no role for our purposes. We will therefore effectively work with the six-dimensional metric
\be
ds^2_\mt{BH} = \left( \frac{r}{R} \right)^{3/2} \left( -f dt^2 + dx_{(3)}^2 + dy^2 \right) +
\left( \frac{R}{r} \right)^{3/2} \frac{dr^2}{f} \,.
\label{six}
\ee

In addition to the metric \eqn{D4metric}, the solution sourced by D4-branes possesses a non-trivial dilaton field given by
\be
e^\Phi = \left( \frac{r}{R} \right)^{3/4} \,.
\ee
Physically, this is one of the most important differences between the D4-brane solution and the analogous D3-brane solution, for which the dilaton is constant. Recalling that the dilaton is related to the SYM coupling constant $\gym$ and that $r$ is related to the energy scale in the gauge theory, we realise that the running dilaton above merely reflects a running coupling constant in the gauge theory, that is, a non-trivial RG flow. Another reflection of this lack of conformal invariance is the fact that the metric \eqn{six} is not that of anti-de Sitter space even if $r_0=0$. Thus, as promised, we have constructed an example of a non-AdS dual of a non-conformal gauge theory. Note that the dilaton diverges as $r \ra \infty$, \ie in the ultraviolet. This corresponds to the fact that the five-dimensional dual gauge theory on the D4-branes is not renormalisable and hence it must be  UV-completed. We will not need the details of this completion here. Suffice it to say that it is provided by the (2,0) superconformal theory on $\nc$ M5-branes, consistently with the fact the D4-brane solution lifts to the eleven-dimenional solution sourced by M5-branes. 

\begin{figure}
\centering{\epsfxsize=9cm\epsfbox{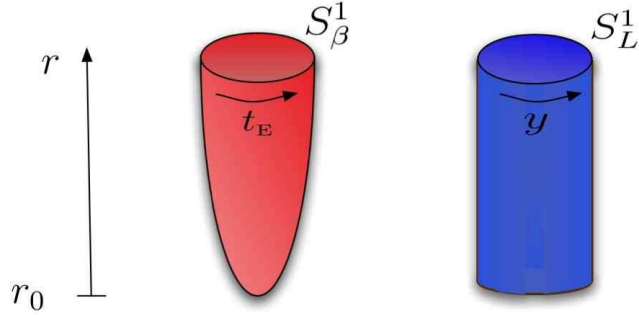}}
\caption{\small The Euclidean continuation of the black hole solution.}
\label{highT}
\end{figure}

The metric \eqn{six} possesses a regular, finite-area horizon at $r=r_0$, so we will refer to it as the black hole solution (hence the subscript in \eqn{six}). In order to determine its Hawking temperature $T$, which we must identify with the gauge theory temperature, we proceed as in section \ref{finite}. We first continue to Euclidean signature via $t \ra i \te$ with the result
\be
ds^2_\mt{E} = \left( \frac{r}{R} \right)^{3/2} \left( f d\te^2 + dx_{(3)}^2 + dy^2 \right) +
\left( \frac{R}{r} \right)^{3/2} \frac{dr^2}{f} \,.
\label{sixeuclid}
\ee
Then we demand regularity at $r=r_0$, which forces us to compactify $\te$ on a circle $S^1_\beta$ of length $\beta=1/T$, \ie $\te \sim \te + \beta$. A simple calculation shows that 
\be
r_0 = \left( \frac{4\pi}{3} \right)^2  \frac{R^3}{\beta^2} \,.
\label{r0}
\ee
A pictorial representation of the Euclidean solution \eqn{sixeuclid}, in which the $x_{(3)}$ directions have been suppressed, is given in fig.~\ref{highT}. 
The key point is that this geometry possesses two compact asymptotic directions, the two circles $S^1_\beta$ and $S^1_L$, parametrised by $\te$ and $y$, respectively. In the interior, the Euclidean time circle shrinks to zero size smoothly at $r=r_0$, so we will refer to this point as the `bottom' of the geometry. In contrast, the  $y$-circle remains non-contractible in the entire Euclidean geometry \eqn{sixeuclid}, or equivalently in the entire region outside the horizon in the Lorentzian geometry 
\eqn{six}. Therefore one must choose whether to impose periodic or antiperiodic boundary conditions for the spacetime fermionic  fields of string theory around 
$S^1_L$. In order words, one must choose a spin structure. This choice is dual to the choice of boundary conditions for the gauge theory fermions around the $y$-direction, so for our purposes we must choose antiperiodic boundary conditions. Note that the boundary conditions around the Euclidean time circle cannot be freely chosen: Because this circle is contractible, regularity at $r=r_0$ requires that fermions be antiperiodic around it. This is consistent with the prescription in thermal field theory according to which fermions must be antiperiodic  around the Euclidean time direction.

\begin{figure}
\centering{\epsfxsize=9cm\epsfbox{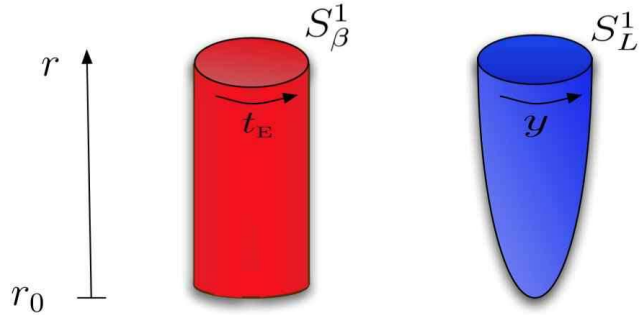}}
\caption{\small The Euclidean continuation of the AdS-soliton.}
\label{lowT}
\end{figure}

Now a simple but far-reaching observation can be made: The metric\footnote{Appropriately combined with the four-sphere factor in eqn.~\eqn{D4metric}.}
\be
ds^2_\mt{E} = \left( \frac{r}{R} \right)^{3/2} \left(  d\te^2 + dx_{(3)}^2 + f dy^2 \right) +
\left( \frac{R}{r} \right)^{3/2} \frac{dr^2}{f}
\label{sixads}
\ee
is also a solution of the supergravity equations.
This is obvious because this metric is obtained from \eqn{sixeuclid} by a simple relabelling of the coordinates $\te$ and $y$. However, in this case it is the $S^1_L$ circle that shrinks to zero size at $r=r_0$, as displayed in fig.~\ref{lowT}, 
so regularity now fixes $r_0$ in terms of $L$ through 
\be
r_0 = \left( \frac{4\pi}{3} \right)^2  \frac{R^3}{L^2} \,,
\ee
which is of course the same as \eqn{r0} with $\beta$ replaced by $L$. 
Despite the simple relation between the Euclidean metrics \eqn{sixeuclid} and 
\eqn{sixads}, the Lorentzian continuation of \eqn{sixads},
\be
ds^2_\mt{soliton} = 
\left( \frac{r}{R} \right)^{3/2} \left(  -dt^2 + dx_{(3)}^2 + f dy^2 \right) +
\left( \frac{R}{r} \right)^{3/2} \frac{dr^2}{f} \,,
\label{sixsoliton}
\ee
differs dramatically from the black hole metric \eqn{six}: The metric \eqn{sixsoliton} describes a completely smooth, horizon-free, Lorentzian spacetime. Following the nomenclature of \cite{HM}, we will refer to this solution as the `AdS-soliton'. A constant-time slice of this spacetime closes off smoothly at $r=r_0$, at which point the spacelike circle parametrised by $y$ shrinks to zero size. Topologically, the $r,y$ coordinates parametrise a plane. In particular, this means that regularity requires that spacetime fermions be antiperiodic around $S^1_L$.

\subsection{The Hawking-Page phase transition}
We are thus confronted with the fact that there exist two candidates for the classical geometry dual to the theory on the D4-branes. This means that the (Euclidean) string partition function \eqn{S}, which we are approximating by the (Euclidean) supergravity action, possesses two saddle points. The ratio of their contributions to the thermal partition function at a given temperature is 
\be
\frac{e^{-\beta F_1 V}}{e^{-\beta F_2 V}} = e^{-\beta \Delta F \, V} \,,
\ee
where $F_i$ are the free energy densities of the solutions and $V$ is the volume of the space on which the gauge theory lives. Since this volume is infinite, any finite difference in the free energy densities translates into an infinite suppression of one contribution with respect to the other. In other words, in the thermodynamic limit of infinite volume the partition function is completely dominated by the saddle point with the smallest free energy. 

The free energies of the black hole and the AdS-soliton can be calculated explicitly by means of standard Euclidean gravity path integral techniques, according to which one identifies $\beta F \equiv S_\mt{E}$, with $S_\mt{E}$ the Euclidean action of the solution in question. Instead of doing this, here we will reason heuristically as follows. Since the Euclidean solutions \eqn{sixeuclid} and \eqn{sixads} are related by the exchange $\te \leftrightarrow y$, the free energy must be symmetric under the exchange of the corresponding periods, $\beta \leftrightarrow L$, or, equivalently, of their inverses $T \leftrightarrow M$. Thus if a phase transition occurs, this must happen at a temperature $\tdec = M$. (The reason for the subscript will become clear below.) Moreover, we expect that the free energy should be minimised by the solution in which the smallest circle at infinity is the one that shrinks to zero size in the interior, the idea being that changing the size of the circle costs gradient energy. These expectations are confirmed by an explicit calculation: For $T < \tdec$ the free energy is minimised by the AdS-soliton, whereas for $T>\tdec$ the minimum-energy solution is the black hole solution. 
At $T=\tdec$ the free energies are equal and a first-order phase transition occurs. In the context of quantum gravity in anti-de Sitter space this is known as a Hawking-Page phase transition \cite{HP}. We will now argue that from the viewpoint of the dual gauge theory this is a confinement/deconfinement phase transition, hence justifying the subscript for the critical temperature.

\subsection{The confinement/deconfinement transition}
Let us now examine the properties of the two phases of the gauge theory dual to the two geometries  \eqn{six} and \eqn{sixsoliton}. The black hole solution \eqn{six} possesses a horizon and hence a non-zero entropy density. A calculation analogous to the one that led to the result \eqn{d3entropy} shows that this scales as $S \sim \nc^2$, as we would expect for an $SU(\nc)$ theory in a deconfined phase. In contrast, the AdS-soliton geometry has no horizon, which implies that the entropy density vanishes in the classical limit, \ie to leading order in $\nc^2$. Consideration of quantum fluctuations leads to an entropy density of order $S \sim \nc^0$, in accordance with our expectation for an $SU(\nc)$ theory in a confined phase. 

\begin{figure}
\centering{\epsfxsize=15cm\epsfbox{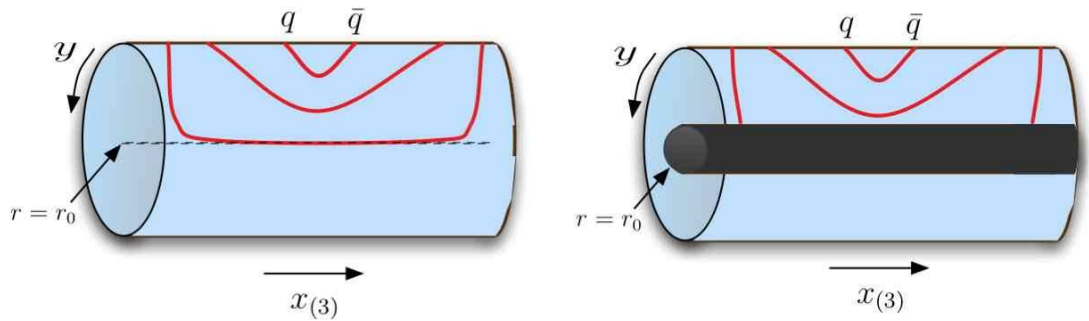}}
\caption{\small String configurations corresponding to a quark-antiquark pair in the AdS-soliton geometry (left) and the black hole geometry (right).}
\label{qqenergy}
\end{figure}

The existence of confinement or lack thereof can also be seen in a more direct way by calculating the energy of a heavy quark-antiquark pair separated in the non-compact gauge theory directions $x_{(3)}$. This is given by the energy of a string ending on the boundary at the location of the quark and the antiquark \cite{wilson}, as illustrated in fig.~\ref{qqenergy}.
In the AdS-soliton geometry the main contribution to the energy of a widely separated quark-antiquark pair comes from the bottom of the geometry and grows linearly with the separation. This leads to a linear, confining potential 
$V_{q\bar{q}} \propto \Delta x$.
In contrast, in the black hole geometry the string can break into two pieces, each of which can fall through the horizon. When this happens the energy becomes independent of the quark-antiquark distance, leading to a non-confining potential 
$V_{q\bar{q}} \sim \mbox{const.}$

The confined and the deconfined phases are also distinguished by their spectra. Gauge theory physical states of definite mass correspond in the gravity description to regular, normalisable modes of the string fields of the form\footnote{More generally, dependence on $y$  and the coordinates on the $S^4$ is also possible.}
$\Phi \sim h(r) e^{ipx}$, with $p^2 = -m^2$. For example, for the dilaton field, which is dual to the operator $\mbox{Tr} F^2$, these modes correspond to glueball states. 
Modes of this type are precisely those used in the construction of the Hilbert space of states on the string side, so the above identification is natural in view of the equivalence between the two theories, which in particular implies an isomorphism between the two Hilbert spaces ${\cal H}_\mt{gauge} \simeq {\cal H}_\mt{string}$. Moreover, it is not difficult to show that the existence of a discrete set of modes of the type above implies, for example, that the two-point function of the dual operator contains a discrete set of poles at $p_i^2=-m_i^2$ (see, for example, \cite{mink,quasi}). In the case of the dilaton, this means that the correlator $\langle \mbox{Tr} F^2 (p) \mbox{Tr} F^2 (-p) \rangle$ has poles at the location of physical glueball states. 

In the AdS-soliton geometry, the spectrum of normalisable modes is discrete and possesses a mass gap. The first property is easy to understand on general grounds. The linearised wave equation for a given mode admits two independent solutions at infinity, and only for a discrete set of values of $p^2$ does the normalisable solution evolve  in the interior to a  solution that is regular at $r=r_0$. A more detailed analysis of the wave equation then shows that this is not possible for $p^2=0$, thus leading to a mass gap \cite{wittenD4}. This confinement scale, which we may call $\lqcd$, turns out to be $\lqcd \sim M$. This is an illustration of a much more general fact that we will discuss in the last section: Within the supergravity approximation, the scale of strong coupling dynamics cannot be decoupled from that of the additional modes in the theory, in this case the Kaluza-Klein modes associated to the fifth dimension.

In the black hole geometry the situation is slightly more subtle. Because of the presence of the horizon, in this case no regularity condition in the interior is needed. 
Consequently, the spectrum of normalisable modes in this phase is continuous and gapless, as one might have expected for a deconfined phase. This statement can be refined by studying the spectral functions of operators with the same quantum numbers as the states of interest. For glueball states, for example, the spectral function in question is 
\be
\chi(p) = 2 \mbox{Im} \,\, i \int d^{4} x \, e^{-i p  x} \, \Theta (t) 
\langle [ \mbox{Tr} F^2 (x), \mbox{Tr} F^2 (0) ] \rangle \,.
\ee

The spectral function encodes the density of states with a given momentum $p$. In the case of a discrete spectrum, as in the AdS-soliton geometry, the spectral function reduces to a sum of delta-funtions located at $p_i^2=-m_i^2$. For a weakly coupled plasma whose excitations can be understood in terms of well defined quasi-particles one would expect the spectral function to exhibit high and narrow peaks. The location of a peak corresponds to the mass of the associated quasi-particle, whereas its narrow width is related to the long lifetime of the quasi-particle. It is remarkable that strongly coupled plasmas with a gravity dual generically exhibit no peaks of this type, but only peaks associated to hydrodynamic modes and peaks whose width is comparable to their height \cite{spectral,spectral2}.\footnote{The location and width of these peaks is related on the gravity side to the real and imaginary parts of so-called quasi-normal modes \cite{mink,quasi}, \ie normalisable modes that satisfy an incoming or outgoing boundary condition at the horizon.} This means that these plasmas behave as strongly coupled liquids whose excitations generically may not be described in terms of well defined quasi-particles, but  possibly only in terms of collective modes.

\section{Conclusions}
In the last section we saw a hint of a generic property of the gauge/gravity correspondence, namely the fact that, because of its asymptotic freedom, a theory like QCD or pure gluodynamics cannot be entirely described within the supergravity approximation. Let us examine the reason for this more closely. 

Suppose we start with a theory whose gravity dual we understand. This could be, for example, the four-dimensional $\caln =4$ SYM theory or the theory on the D4-branes we studied in the last section. Imagine we then deform this theory by giving a mass of order $M$ to all fields except to the gluons. In the case of D4-branes, this can be done as above by compactifying the theory on a circle with supersymmetry-breaking  boundary conditions, which also renders the theory effectively four-dimensional at low energies. In the case of the $\caln =4$ theory, we could simply introduce explicit masses for the scalars and the fermions. In any case, we hope that the theory at sufficiently low energies reduces to pure Yang-Mills theory, which ought to develop a confining scale $\lqcd$. Generically, corrections to the pure Yang-Mills dynamics will be suppressed by powers of  $\lqcd/M$, so we would like to study the regime in which these two scales decouple, $\lqcd \ll M$. However, these scales are related through
\be
\lqcd \sim M \exp \left( -\frac{c}{\gym^2 (M) \nc} \right) \,,
\ee
where $c$ is a positive constant and $\gym(M)$ is the Yang-Mills coupling constant evaluated at the scale $M$. This equation is just the statement that the coupling constant runs logarithmically below the scale $M$. It follows that decoupling $\lqcd$ from $M$ requires that the 't Hooft coupling $\lambda$ be very small at the scale $M$, \ie 
$\lambda \ll 1$. This of course is just a reflection of the fact that pure Yang-Mills is asymptotically free. 
On the other hand, we know that the supergravity approximation is reliable in the opposite regime, \ie when $\lambda \gg 1$. In this regime $\lqcd \sim M$ and the effect of the unwanted fields is not suppressed. 

Since at present we do not know how to systematically go beyond the supergravity approximation in the type of string backgrounds of interest, the above analysis shows that certain quantitative features of QCD, for example its zero-temperature mass spectrum, cannot be reliably studied. However, this does not mean that certain properties derived from supergravity cannot be universal or robust enough to apply to   
QCD, at least in certain regimes. The viscosity/entropy ratio explained in section 
\ref{finite}, or the existence of universal phase transitions for fundamental matter \cite{us}, are examples of such properties. Understanding more precisely what properties can be studied within supergravity, to what regimes of QCD they can be applied, and how to quantify the accuracy of the approximation, are important challenges for the near future. On a longer term we may be able to go beyond the supergravity approximation. This may make it possible to realise the long-standing hope of solving the large-$\nc$ limit of QCD through a string dual.

\section*{Acknowledgments}

\noindent
I am grateful to the organisers of the RTN Winter School on ``Strings, Supergravity and Gauge Theories" at CERN on January 15-19, 2007, for the opportunity to present these lectures, and to them and to all the participants for providing an stimulating atmosphere with many challenging questions. I would also like to thank the Theory Division at CERN for hospitality, as well as Sean Hartnoll and Don Marolf for discussions and useful comments on this manuscript. I am supported by NSF grant no PHY-0555669.

\end{document}